\documentclass{tufte-handout}

\usepackage{amsmath}

\usepackage{url}

\usepackage{graphicx}
\setkeys{Gin}{width=\linewidth,totalheight=\textheight,keepaspectratio}
\graphicspath{{graphics/}}

\usepackage{calc}
\usepackage{enumitem}
\newlength\widest

\usepackage{booktabs}

\usepackage{units}

\usepackage{fancyvrb}
\fvset{fontsize=\normalsize}

\usepackage{multicol}

\usepackage{lipsum}

\usepackage{hyperref} % provides \url{}
\usepackage{enumitem}
\setlist{nosep}

\title{Large Language Models Humanize Technology}
\author{Pratyush Kumar, Microsoft Research \& AI4Bharat}
\begin{document}

\maketitle% this prints the handout title, author, and date

\begin{marginfigure}
\includegraphics{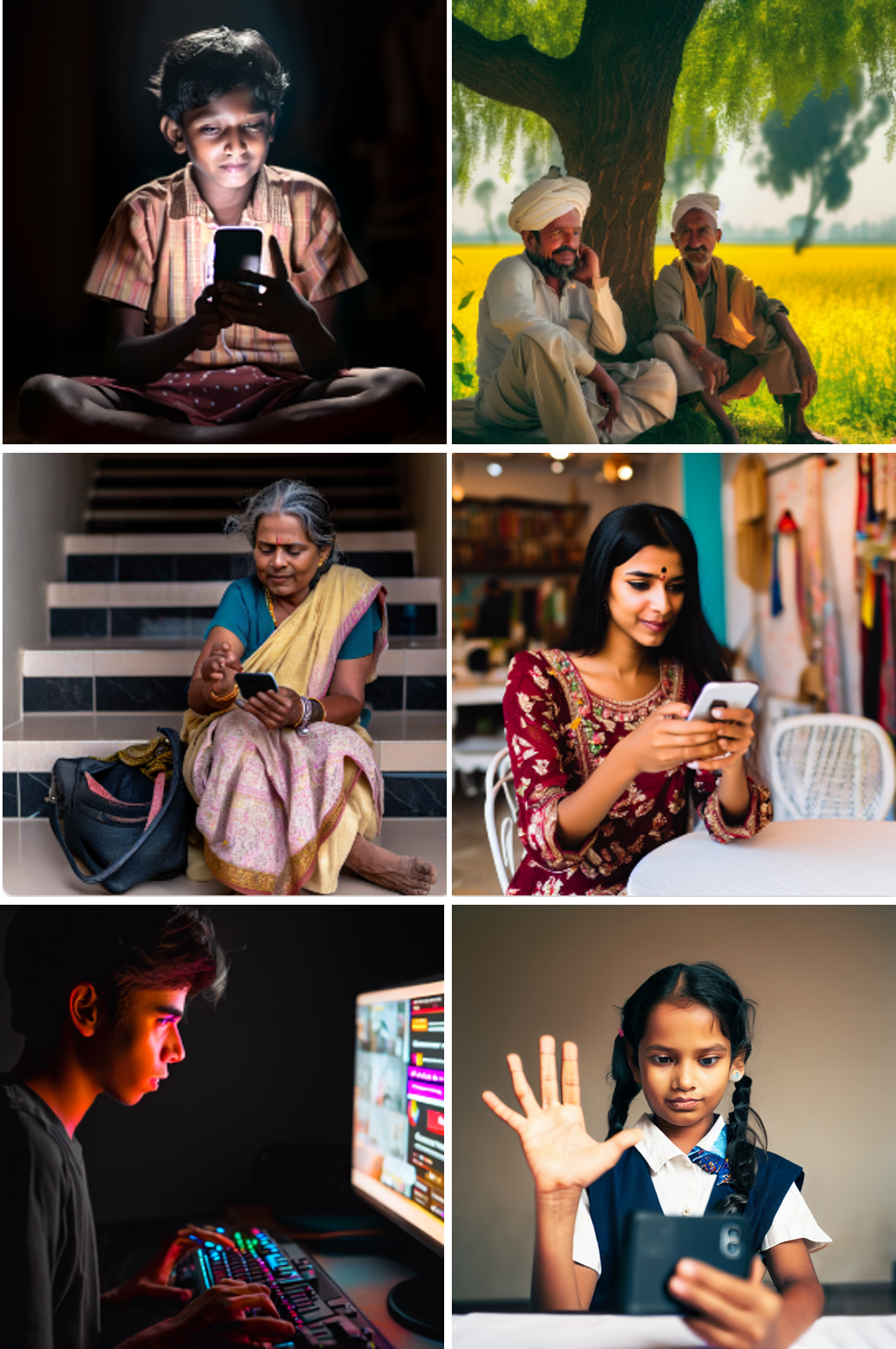}
\caption{Illustrations of fictional people, mentioned in the text, who would today face mechanizing bottlenecks with technology which can be addressed with emergent abilities of LLMs.
All pictures are generated with DALL-E 2.}
\end{marginfigure}

\begin{abstract}

% Large Language Models (LLMs) have rapidly progressed in the last few months and weeks, that too in the public eye.
% This has sparked concern on aligning these models to human values, their impact on labour markets, and if and how to regulate further research and development. 
% However, missing from this discourse is an articulation on the imperative to diffuse the societal benefits of LLMs widely. 
% To qualify this societal benefit, we assert that LLMs exhibit emergent abilities to \textit{humanize technology}, far more effectively than previous technologies and for people across language, occupation, and accessibility divides.
% We argue that they do so by addressing three mechanizing bottlenecks in today's computing technologies: creating diverse and accessible content, learning complex digital tools, and personalizing machine learning algorithms. 
% We take a case-based approach and illustrate each bottleneck with two examples where current technology imposes bottlenecks which LLMs demonstrate ability to address.
% Given this opportunity to humanize technology widely, we advocate for more widespread understanding of LLMs, tools and methods to simplify use of LLMs, and cross-cutting institutional capacity for co-invention.

Large Language Models (LLMs) have made rapid progress in recent months and weeks, garnering significant public attention. This has sparked concerns about aligning these models with human values, their impact on labor markets, and the potential need for regulation in further research and development. However, the discourse often lacks a focus on the imperative to widely diffuse the societal benefits of LLMs.
To qualify this societal benefit, we assert that LLMs exhibit emergent abilities to humanize technology more effectively than previous technologies, and for people across language, occupation, and accessibility divides. We argue that they do so by addressing three mechanizing bottlenecks in today's computing technologies: creating diverse and accessible content, learning complex digital tools, and personalizing machine learning algorithms.
We adopt a case-based approach and illustrate each bottleneck with two examples where current technology imposes bottlenecks that LLMs demonstrate the ability to address. Given this opportunity to humanize technology widely, we advocate for more widespread understanding of LLMs, tools and methods to simplify use of LLMs, and cross-cutting institutional capacity.

\end{abstract}

\section{Introduction}

Large Language Models (LLMs), such as the GPT series, have rapidly progressed in the last few years, with accelerating advances in the recent past.
Key to this trend has been the consistent improvement observed when scaling both the Transformer-based models and the text datasets on which they are trained.\cite{radford2019language}
At the scale of trillion-sized models and petabyte-sized training data, LLMs excel at tasks such as translation, summarization, creative writing, and code generation\cite{liang2022holistic}, outperform humans on many professional examinations\cite{openai2023gpt4}, and broadly exhibit signs of general intelligence.\cite{bubeck2023sparks}
These are referred to as \textit{emergent abilities} as they are unexpected abilities not demonstrated by smaller models.\cite{wei2022emergent}
% Previously, solving such tasks required domain-specific datasets and expert-driven AI model design and training.
Given these broad-based abilities, LLMs exhibit properties of general purpose technologies\cite{bresnahan1995general} and are expected to have transformative impact across domains. 

With the launch and popular adoption of tools such as Github Copilot and ChatGPT, the capabilities of LLMs are widely recognized.
This has sparked a public debate on the potential risks including discrimination implicit in the model, misinformation due to hallucination and reasoning errors, misuse by bad actors, and rapid socio-economic changes.\cite{weidinger2022taxonomy}
Research continues on making LLMs more reliable by aligning them with human feedback\cite{ouyang2022training}, with early observations that larger models can be made more robust.\cite{openai2023gpt4}
Socio-economically, there is increasing concern of job losses due to automation.\cite{eloundou2023gpts}
Some posit existential risks due to runaway consequences of AI surpassing human intelligence, and call for a pause on research and development on giant models.\footnote{\url{https://futureoflife.org/open-letter/pause-giant-ai-experiments/}}
Overall, the study of risks of LLMs and mitigation strategies is scientifically nascent, especially given the limited access to giant LLMs within academia.

However, missing from the discourse on responsible AI is attention on societal good realisable with LLMs and the imperative of equitable diffusion to bridge digital divides with long-lasting effects.\cite{van2020digital}
Parallels can be drawn to the societal impact of mobile technologies, which in the last decade has enabled billions of people with improved outcomes across education, health, and finance.\cite{rotondi2020leveraging}
In comparison, LLMs may have a faster and more substantial impact given that digital technologies are known to accelerate across generations.\cite{comin2010exploration}
In this essay, we assert that an emergent ability of LLMs is to \textit{humanize technology}, a term that we use to imply making technology more human-centered, accessible, and responsive to human needs and values.\cite{norman2013design}
We make this argument by calling out three mechanizing bottlenecks in current technology, namely, creating diverse and accessible content, learning complex digital tools, and personalizing machine learning algorithms.
These bottlenecks reinforce digital divides and sustain dark patterns.\cite{eyal2014hooked}
For each bottleneck, we illustrate two examples where LLMs demonstrate ability to address the bottlenecks and humanize technology for people across language, occupation, and accessibility divides.\footnote{See Figure~1 for a collage of fictional characters considered for these examples}
We take this 'case-based' approach to share preliminary enthusiasm on what is already possible, and hope for a more grounded theory\cite{glaser2017discovery} to emerge over time.
Our primary hypothesis in this essay is that LLMs of today demonstrate ability to humanize technology in unexpected and diverse ways far exceeding the capabilities of earlier technologies.
Further, the generality in the intelligence of these models parallels their generality in applicability across contexts and societies, as a  technology ``flattener.''\cite{friedman2005world}

We also propose a three-point agenda for effective diffusion of this technology.\cite{rogers2010diffusion}
First, we need wider understanding of LLMs' abilities, especially since much of them are  unknown and only elicited through culturally rich and creative usage. 
Second, we need to engineer tools and methods to enable roles of agency and responsibility for human facilitators in building applications with LLMs.
Third, we need institutional capacity at the intersection of academia, government, and technology companies for co-invention\cite{bresnahan1996technical} around the core technology of LLMs.
A relevant example is the success of developing and scaling Digital Public Goods (DPGs).\cite{raghavan2019india}

% LLMs as a general purpose technology are amenable to diffusion based on the five criteria laid down by Everett Rogers\cite{rogers2010diffusion}.
% For effective diffusion, we argue for complementary innovation around the core technology of LLMs to mitigate current risks of bias and misinformation and to create roles of agency and responsibility for human facilitators.
% We also make the case for institutional capacity spanning science, state, and startups to accelerate innovation and to bring greater shared understanding of the LLMs and their impact.

\section{Creating Diverse and Accessible Content}
While digital experiences enable scale, they often dislocate the creation and consumption of content.
For instance, in e-learning, a teacher creates content explaining a topic which is viewed by students often not in direct touch with the teacher. 
Or in e-governance, a public official lists details of a government scheme on a website which are then accessed by citizens often not in direct touch with the official. 
This dislocation requires that the creation process anticipate and design for a wide range of accessibility and informational needs of the users. 
The teacher is required to create content that various students would find engaging, while the official is required to create content across languages and anticipate various citizens' questions.
This need for generating diverse and accessible content is a mechanizing bottleneck that can lower engagement or create informational asymmetries. 

\subsection{Example 1: Student learns with creative play}

\begin{marginfigure}
\includegraphics{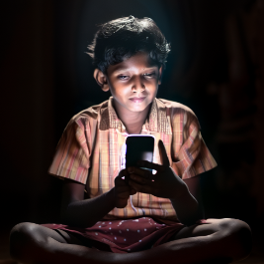}
\caption{A high school student asks an LLM to generate a discussion between Hanuman and Thanos. Image generated by DALL-E 2."}
\end{marginfigure}

In the influential work on \textit{Play, Dreams and Imitation in Childhood}, psychologist Jean Piaget writes\cite{piaget2013play}, ``The great principle of the universe of play is, in fact, that the child transforms the world in order to adapt it to his own level of assimilation and to his own requirements.''
Eight decades after the book by Piaget, LLMs today may enable this level of personalized play within the context of learning.
Consider a high school student Ravi asking an LLM: "If Hanuman\footnote{A monkey God from the Indian epic 'Ramayana' who is known for his strength, devotion, and selfless service.} and Thanos\footnote{A character from the Marvel Universe who is driven by a misplaced motive to eradicate half of all life in the cosmos.} were to meet, what would they talk about?".
Though the two characters in question span vast differences in cultural, temporal, and creative origins, they may coincide in Ravi's lived experience and pique his curiosity. 
It is unrealistic for teachers to anticipate such coincidences and generate content for them. 
However, GPT-4 is able to answer this question with fluency, accuracy, and insight drawing upon culturally nuanced data,\footnote{See Exhibit in the next page} as an ability emergent with no close precedent in any earlier AI models.
In this setting, LLMs humanize Ravi's interaction with the learning content by suffusing it with agency and play\footnote{Admittedly, there can be far richer imaginations of play between a student and a learner, for instance co-writing a story as demonstrated in Khanamigo \url{https://www.khanacademy.org/khan-labs}}, without being bottle-necked by the static content created by his teacher.
The role of the teacher then is transformed to creating scaffolds\cite{wood1976role} on top of LLMs to guide the students' experience based on pedagogic insight.
For instance, the teacher could provide the instruction to pick a word from GPT's response that the student is unfamiliar with and ask a follow-up question.\footnote{See Exhibit on the follow-up question on Dharma}
And the goal of education itself becomes one of cultivating curiosity, of asking the right questions and exploring the world around us and the world that could be.

\begin{figure*}
\includegraphics{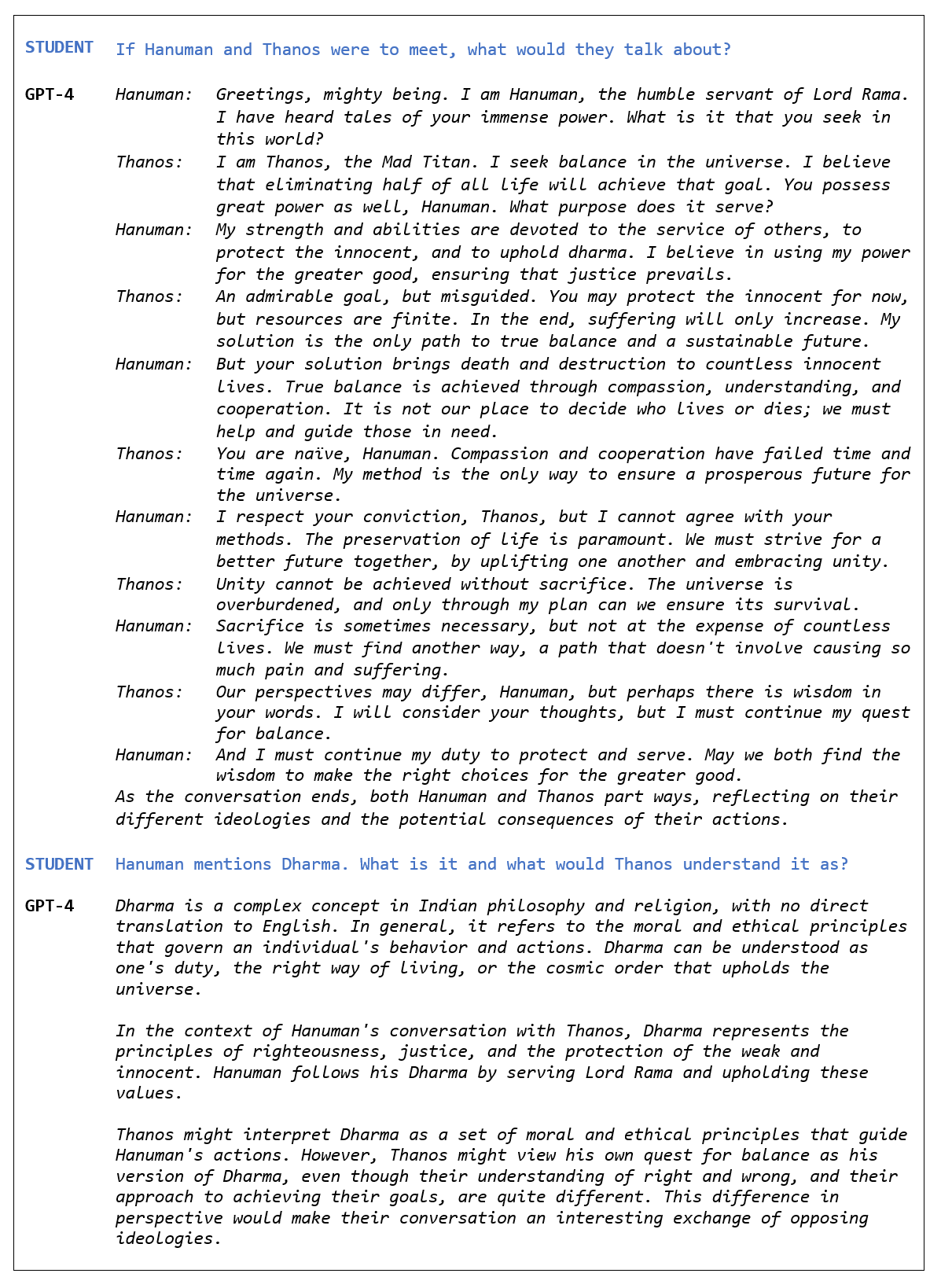}
\caption{Exhibit for Example 1: Conversation between Hanuman and Thanos}
\end{figure*}
\pagebreak

\subsection{Example 2: Farmer understands a government scheme}

\begin{marginfigure}
\includegraphics{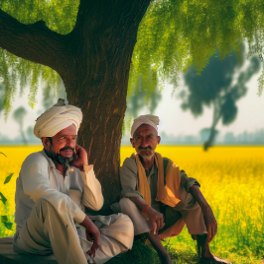}
\caption{A farmer speaks in Punjabi to an LLM powered voice call system to answer custom queries reasoned over an English document. Image generated by DALL-E 2.}
\end{marginfigure}

Digital divide is identified as a key challenge in effective e-governance.\cite{belanger2009impact}
Mark Warschauer specifically calls out language\cite{warschauer2004technology}, ``Language is a central factor in the digital divide, shaping not only people's access to technology but also how they use it, and what kind of information and communication resources they are able to draw on.''
Governments in linguistically diverse nations like India find it hard to create accessible portals of information. 

In this context, consider a LLM-powered voice bot that enables a farmer Jaswinder to dial in and conversationally interact in Punjabi on rights and entitlements.
The bot transcribes Jaswinder's speech using speech recognition models\footnote{Speech recognition and synthesis models for various Indian languages are now available in the open-source through the Bhashini project: \url{https://bhashini.gov.in}} and then uses LLMs to respond to Punjabi queries using English documents.
For instance, Jaswinder may ask whether he is eligible based on his land size of 4 acres. 
In this case, the bot\footnote{see Exhibit in the next page} is able to reason with unit conversion and reply that Jaswinder is eligible given that 4 acres is smaller than the allowed limit of 2 hectares.\footnote{Incidentally, this specific example requiring conversion on land area units used in India did not work with the earlier generation of the GPT models even in English, suggesting greater cultural inclusivity with models trained on larger amounts of data.} 
Further, the scheme details may specify the common use-case of applying through the help of village level entrepreneurs, but Jaswinder may be digitally able to apply himself. 
He may ask in Punjabi about how to do so, and the bot may reply with the procedure to apply at a web portal.\footnote{See Exhibit in the next page}
In this setting, LLMs humanize Jaswinder's interaction with the government.
The role of the officials is transformed from the mechanizing work of updating multilingual websites to understanding and responding to citizen needs.
To provide every citizen such conversational experiences, the government may invest in building open and accurate \textit{interface AI models} such as speech-to-text and text-to-speech across languages.
Thereby, e-governance may advance to its true goals of ``transparency, enhanced citizen engagement, and more efficient government operations'' as outlined by Richard Heeks over two decades ago.\cite{heeks2001understanding}

\newthought{The two examples above} draw upon distinct emergent abilities of LLMs. 
The student's interaction depends on encoding diverse data during pretraining, while the farmer's interaction depends on cross-lingual reasoning on custom data.
Though our limited experiments were successful, formal evaluation is required to benchmark LLMs on eliciting diverse cultural knowledge, working across languages, and reasoning on various domains of information. 
For instance, for low-resource languages with smaller representation in pretraining, the accuracy is likely to be lower and needs quantified evaluation. 
In addition, co-invention is required to easily design and robustly implement guardrails for allowable interactions.

\begin{figure*}
\includegraphics{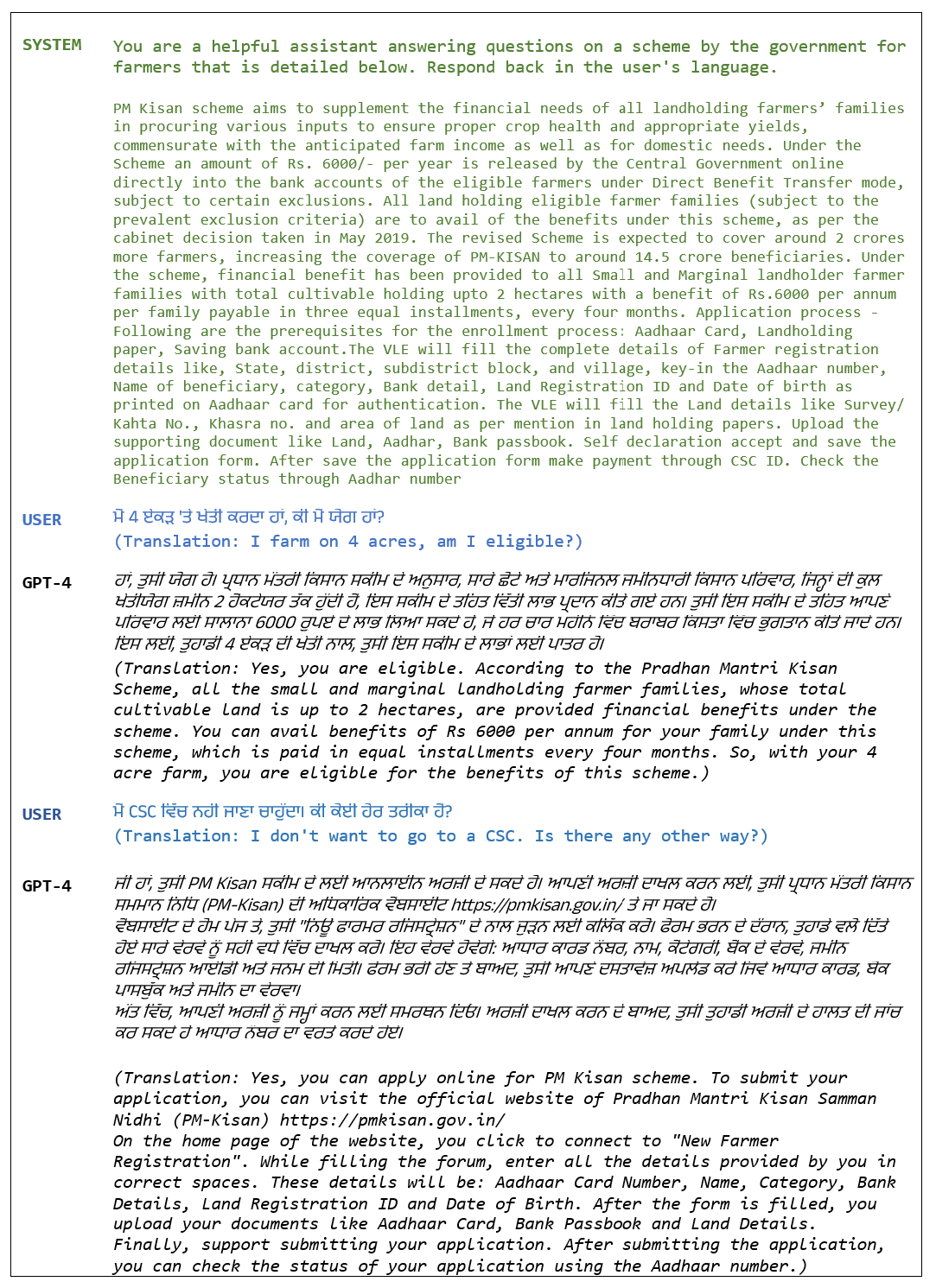}
\caption{Exhibit for Example 2: Farmer speaking in Punjabi and getting queries answered against an English document}
\end{figure*}

\section{Learning Complex Digital Tools}
Dutch sociologist Jan van Dijk in his longitudinal study of digital divide\cite{van2020digital}, identified four layers of digital divide, namely, motivational access, material access, skills access, and usage access. 
Critically, Dijk emphasizes skills: even if someone has the motivation, the resources, and the use-case, if they did not have the digital skills necessary, they would not be able to benefit from digital technologies. 
It is useful to understand an instance of this in relation to smartphones.
Though Indians represent over 17\% of internet users\footnote{\url{https://en.wikipedia.org/wiki/List_of_countries_by_number_of_Internet_users}}, only 3\% of the apps are designed by India developers.\footnote{\url{https://42matters.com/india-app-market-statistics}}
Also, based on statistics of app downloads\footnote{\url{https://www.similarweb.com/apps/top/google/app-index/in/all/top-free/}}, the top smartphone apps in India are primarily focused on communication and entertainment, with very few apps in the category of productivity.
These statistics suggest that programming environments for apps and productivity apps are complex digital tools whose learning imposes a bottleneck.
Poor English language skills are a significant contributor to this bottleneck.\cite{becker2019parlez}
And this bottleneck alienates people from learning computational thinking, which as Jeannette Wing\cite{wing2008computational} wrote in 2008, is a skill that ``influences everyone in every field of endeavour.''

\subsection{Example 3: Domestic help does her finances on a chat app}

\begin{marginfigure}
\includegraphics{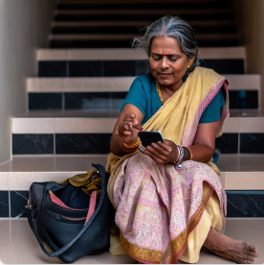}
\caption{A domestic help converses with an LLM bot and do her finances without knowing how to operate a spreadsheet app. Image generated by DALL-E 2."}
\end{marginfigure}

Consider a domestic help, Padma, who collects her monthly wages in cash from her employers. 
With the rapid digital transformation of financial services in India, Padma is able to access banking facilities like deposits and digital payments from her smartphone. 
But Padma would like to do more, to maintain her own log of her income, expenses, and advances she has taken from employers, and plan financially towards a family purchase in the traditional festival of Deepavali.
A spreadsheet app, such as Microsoft Excel, is well suited and widely used for such a task.
However, since Padma speaks only Kannada, has never used a computer, or taken any formal training in computational thinking, she will face significant challenges in learning and using Excel. 
This is a mechanizing bottleneck imposed by technology today. 
In this context, consider an LLM-powered bot that enables Padma to enter her income, her current savings, and create a plan for how much to save to reach a financial goal. 
Such a conversation in Kannada works fairly well with GPT-4.\footnote{see Exhibit in the next page}
A more robust version of this could use a spreadsheet application as a plugin to GPT.
Thereby, LLMs literally interface Padma over the digital divide onto complex tools like spreadsheets.

\begin{figure*}
\includegraphics{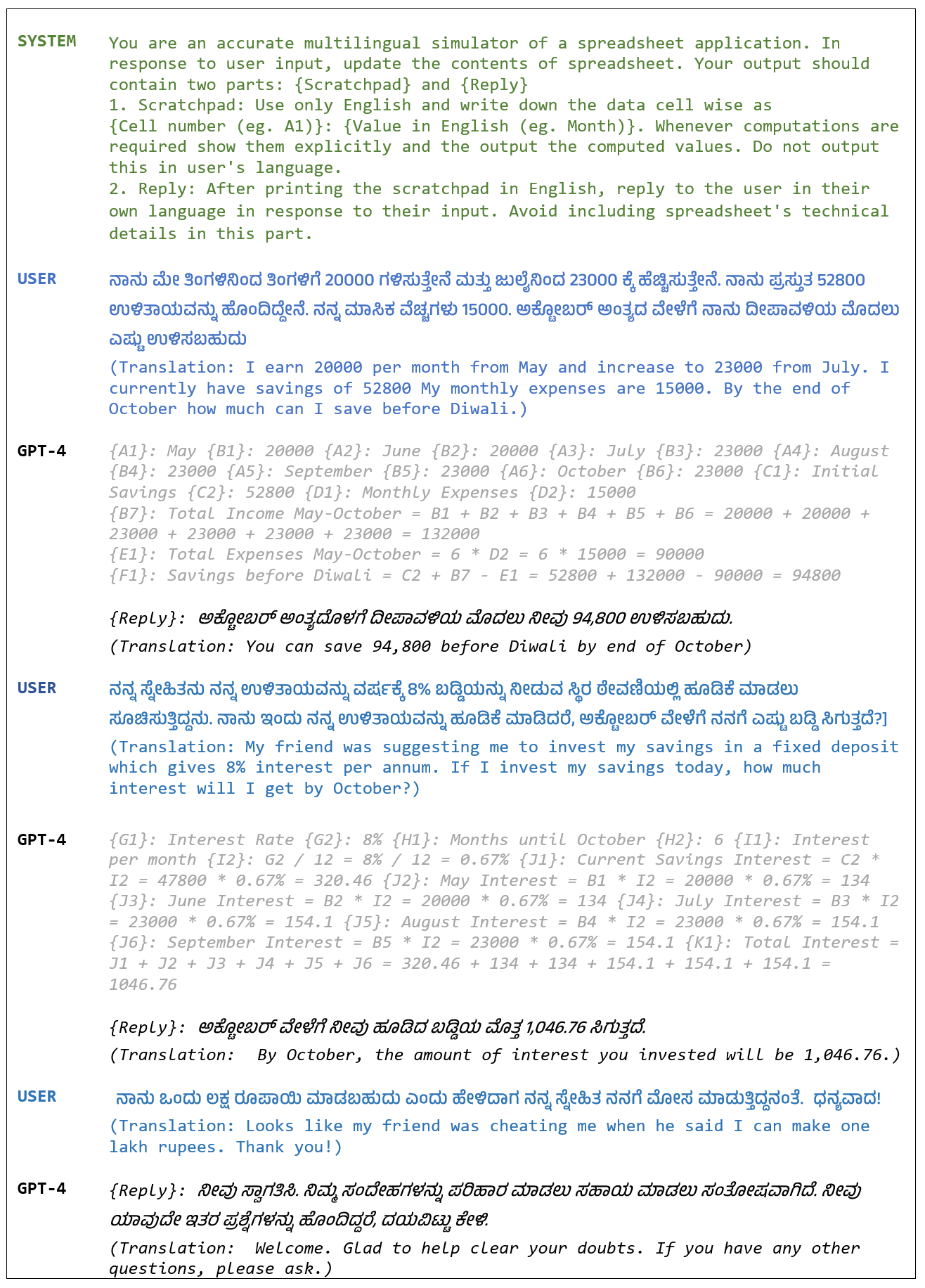}
\caption{Exhibit for Example 3: Simulation of an Excel sheet for helping a user do their finances}
\end{figure*}
~
\pagebreak

\subsection{Example 4: Non-programmer writes code to draw a Rangoli}

\begin{marginfigure}
\includegraphics{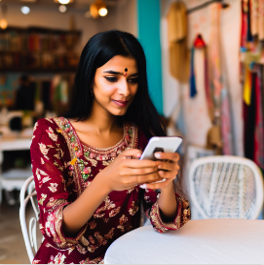}
\caption{A non-programmer prompts LLMs to generate code that draw images. Image generated by DALL-E 2."}
\end{marginfigure}

Seymour Papert wrote ``The computer is the Proteus of machines. Its essence is its universality, its power to simulate. Because it can take on a thousand forms and can serve a thousand functions, it can appeal to a thousand tastes.''\cite{papert1980children}
This quote has aged beautifully over the last four decades as use and programming of computers today pervade various aspect of our lives. 
However, only a very small fraction of the world's population is skilled  to program computers, and thus intentionally build digital systems. 
% This digital divide is reinforced by various existing inequities such as access to computers and formal education, and also knowledge of English.
% Can LLMs address this digital divide?
Models such as Codex are being extensively used to enable higher productivity for developers. 
On the flip side, can LLMs enable non-programmers the ability to robustly program?
Specifically, can LLMs enable ``encoding the programmer's contributions in units that correspond to the programmer's intentions'', a goal enlisted by Charles Simonyi much ahead of his time in defining intentional programming in 1995\cite{simonyi1995death}.
We consider this question in the context of a young working mother, Anu, who wants to program a computer to draw diagrams of Rangoli\footnote{From Wikipedia: Rangoli is an art form that originates from the Indian subcontinent, in which patterns are created on the floor or a tabletop using materials such as powdered lime stone, red ochre, dry rice flour, coloured sand, quartz powder, flower petals, and coloured rocks}.
However, as she does not know how to code, she writes her intention in Hindi and is able to ask an LLM to generate LaTeX code which then generates the figure.\footnote{Anu could also use image generation models like DALLE-2 to generate images of Rangoli. Though, in this specific case LaTeX code is probably more suited at generating low-dimensional line art as may be of interest to Anu in drawing her Rangoli. However, more substantially, GPT-4 generating code and then images, provides more control on the process as seen in the Exhibit.}
This works reasonably well with GPT-4\footnote{See Exhibit in the next page} and Anu is also able to change the colour in a follow-up utterance.
This is remarkable on three counts: (a) LLM understands the cultural nuance of Rangoli, (b) LLM generates well-formed LaTeX code which ingeniously draws shapes with braces and curves, and (c) LLM generates Hindi comments which Anu could use to read and begin to understand code over time.  
The last point is an emergent generalization as GPT is unlikely to have been pretrained on code with Hindi comments, and indicates  potential of flattening the differences across languages.

\newthought{In both the above examples}, we depend on the outputs of LLMs being precise, either in terms of arithmetic computations or in generating syntactically correct code. 
In some of our experiments, we found that LLMs made errors in these steps, specifically arithmetic, as has been widely reported.\footnote{\url{https://spectrum.ieee.org/large-language-models-math\#toggle-gdpr}}
However, this may be an transient situation, as LLMs need not directly compute but can invoke tools\cite{schick2023toolformer} such as calculators and program execution engines.\cite{gao2023pal}
Indeed, services such as ChatGPT may work with spreadsheet applications and code compilation tools as 'plugins'\footnote{\url{https://openai.com/blog/chatgpt-plugins}}. 
Seen in this light, LLMs may become accessible interfaces to a range of digital tools across languages and use-cases of end-users.
Crucially, there is opportunity in enabling this simultaneously to improve the productivity of knowledge workers while enabling an entirely new population to become first time users of these tools.

\begin{figure*}
\includegraphics{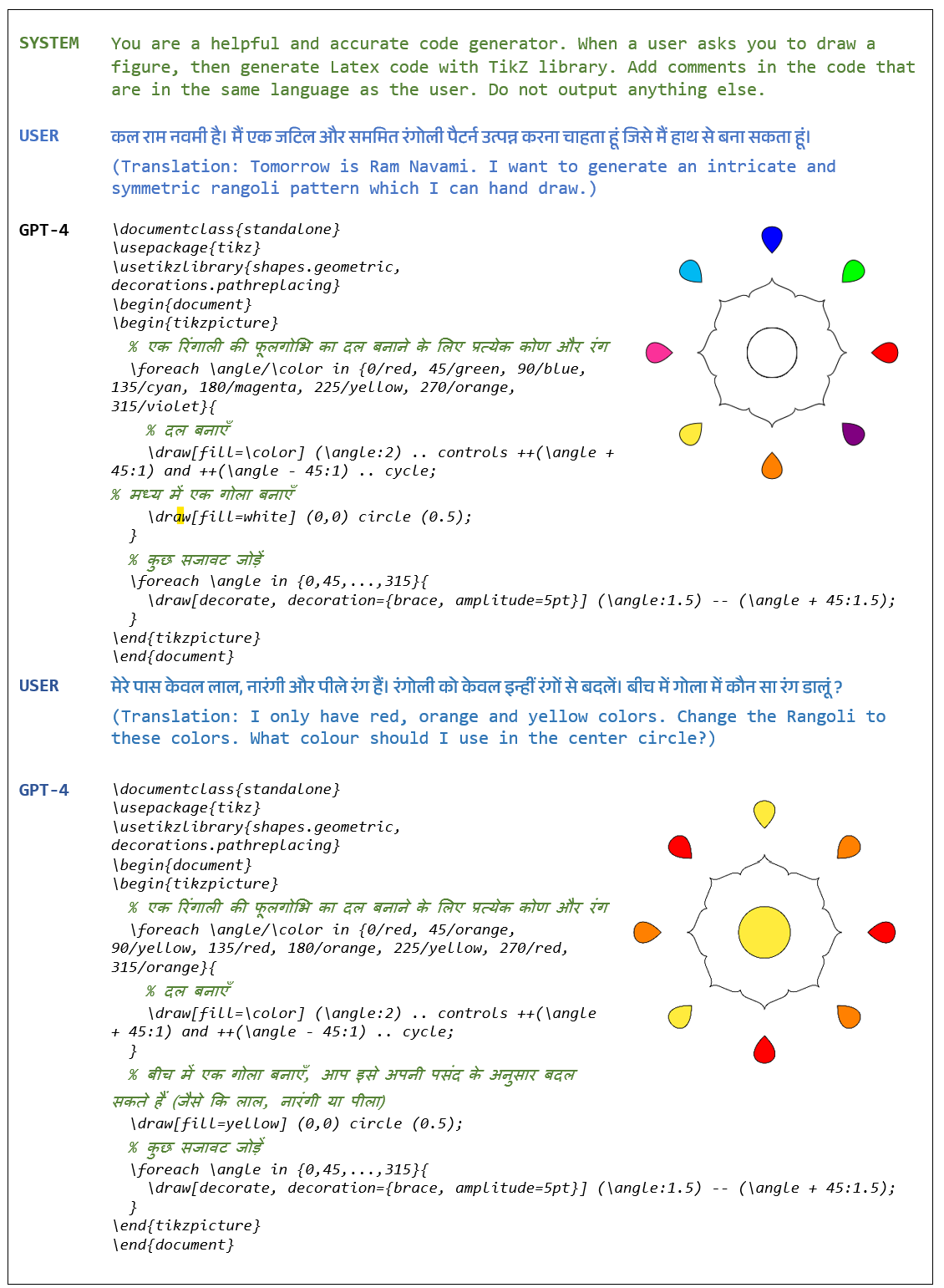}
\caption{Exhibit for Example 4: Intentional programming in Hindi to get generate Rangoli art drawn through LaTeX}
\end{figure*}

\section{Configuration of Machine Learning Algorithms}
% In a way, the examples discussed so far have walked through the phases of evolution of digital technologies in the following order:
% digital technologies enabling long-range accurate communication (farmer's query example),  creation and sharing of data through various mediums including early internet (student's story example), usage of productivity tools such as Microsoft Word (financial planning with spreadsheet example), and programming at scale including over 30 million people developing smartphone apps (programming rangoli example).
The last decade in computing has been dominated by the field of machine learning (ML) wherein computers learn to perform tasks from large amounts of data. 
ML powers much of the internet economy, including the online advertisement industry valued today at over half a trillion dollars\footnote{\url{https://www.statista.com/outlook/dmo/digital-advertising/worldwide}}, and relies on collecting data of consumers and producers through two-sided marketplaces.\cite{rochet2006two}
ML also powers richer human-computer interactions including accurate speech recognition and augmented virtual reality. 
% Each of these ML advances requires large amounts of data, training AI models on powerful compute infrastructure, and then deploying these models efficiently on the cloud.
The process of building ML models is expensive and hence exclusionary as it requires large amounts of data, compute, and AI expertise.
The question to ask now is: Can LLMs enable large population of users with entirely new simplified processes of building personalized intelligence?

\subsection{Example 5: A Teenager Speaks to a Recommendation Algorithm}
\begin{marginfigure}
\includegraphics{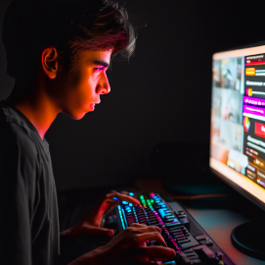}
\caption{A user of a video website builds a personal recommendation engine based on his nuanced preferences. Image generated by DALL-E 2.}
\end{marginfigure}

Recommendation algorithms\cite{narayanan2023understanding} are a dominant kind of machine learning algorithm driving social media platforms such as YouTube, Facebook, and TikTok.
While these platforms have been successful in scaling, their functioning has revealed many issues on the ethics of intentionally designing ``habit forming systems''.\cite{eyal2014hooked}
Longitudinal studies have synthesized evidence on the influence of such systems on depression, anxiety, and psychological distress in adolescents.\cite{keles2020systematic}
While these concern ethical choices made by corporations, the technical challenge remains: Today's recommendation algorithms need large amounts of data collected across users and across platforms.
Can LLMs enable an alternative approach for personalized recommendation algorithms?
Consider a teenager, Ahmed, who often has different preferences from his friends. 
He wants to find the movie to watch next and could use recommendation algorithms on Netflix and Youtube to decide this. 
As an alternative, he has a conversation with GPT-4 explaining his recent experiences watching movies with nuanced details of what he liked and disliked.
GPT-4 can process this information and then recommend movies with detailed reasoning for the recommendation.
This works fairly well\footnote{see Exhibit in the next page}, and enables Ahmed to ask a follow-up question with an added filter.
While this is a demonstration, production grade systems can be built with LLMs grounded on movie databases.
Critically, such systems would not impose dehumanizing bottleneck of tracking and collecting click logs of users across websites, and instead rely on rich engaging conversations.
And these conversations would afford users both interpretability and explainability of the decision made, unlike current opaque systems.

% Recommender systems could be engineered by building LLMs grounded on a domain (e.g., movies) and to make relevant recommendations not by aggregating information across users on all platforms, but by engaging in richer conversations about nuances.
% Further, such systems can afford users such as Ahmed the ability to inspect and selectively edit their profiles over time, thereby returning to them agency and humanity.

\begin{figure*}
\includegraphics{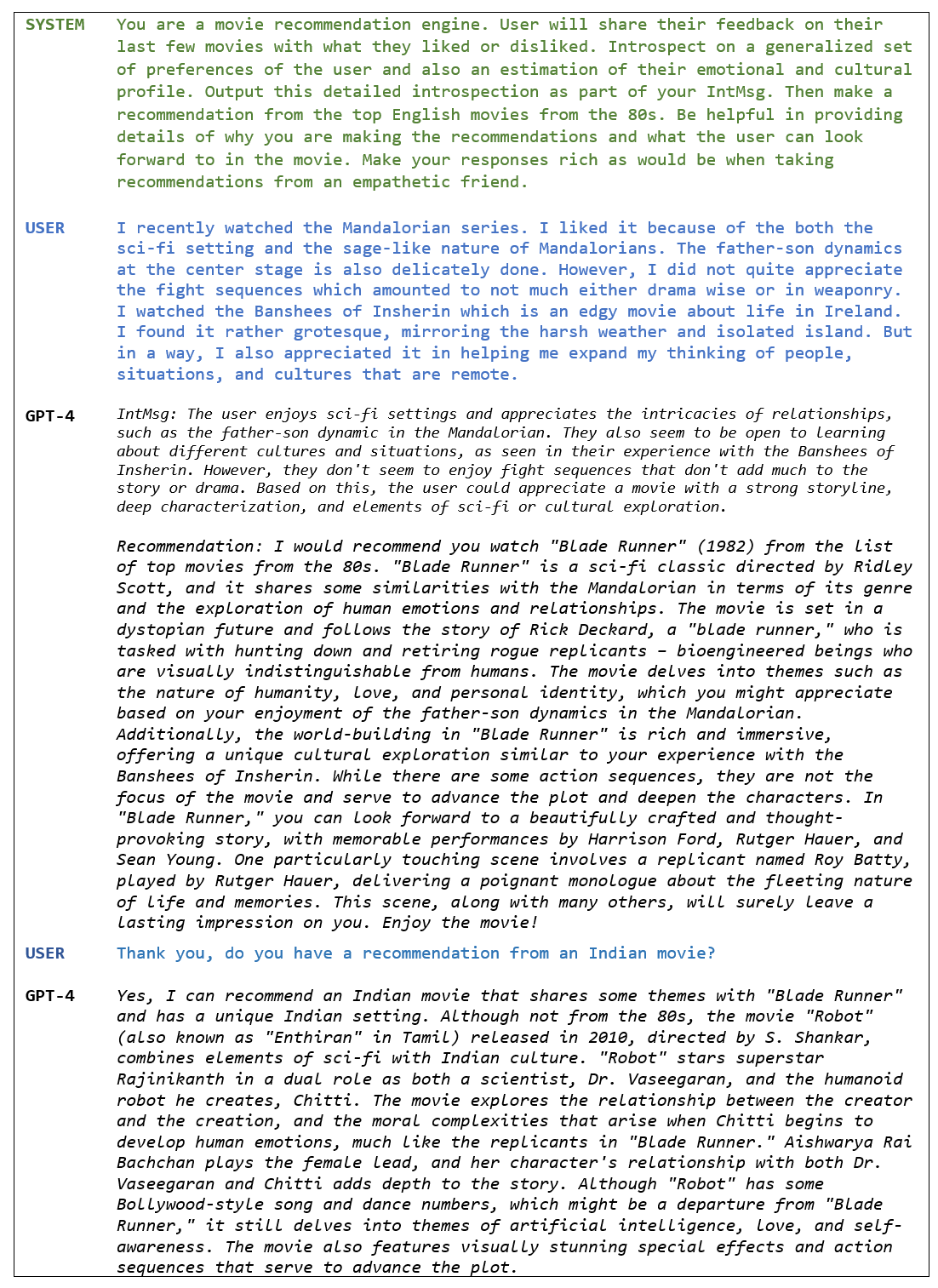}
\caption{Exhibit for Example 5: A personalized recommendation for a movie based on nuanced input by a user}
\end{figure*}

\begin{marginfigure}
\includegraphics{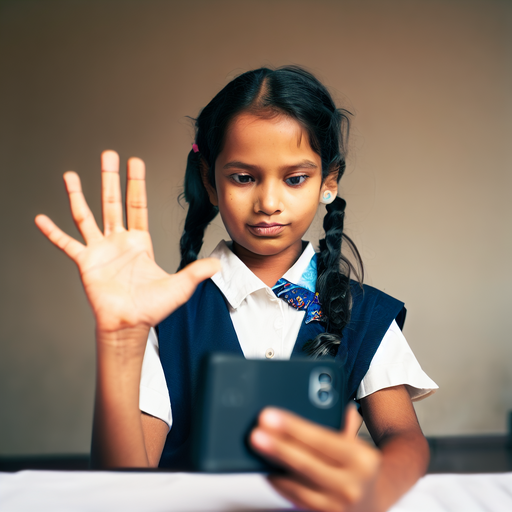}
\caption{A Deaf girl teaches her mobile to recognize her signs. Image generated by DALL-E 2.\protect\\~\protect\\}
\end{marginfigure}
\subsection{A Deaf girl trains a sign language recognition model}
Over 5\% of the world experiences hearing disabilities\footnote{\url{https://www.who.int/news-room/fact-sheets/detail/deafness-and-hearing-loss}}, but are not adequately served by machine learning tools today.
For instance, translation models between natural languages, say Hindi and French, are fairly accurate.
However, models to recognize sign language, across the three hundred sign languages\footnote{\url{https://en.wikipedia.org/wiki/List_of_sign_languages}}, are inaccurate primarily due to the lack of datasets.\cite{bragg2021fate}
This need for creating huge datasets is a mechanizing bottleneck that machine learning algorithms impose.
Consider 10-year old Mary who is Deaf and wants to teach her mobile phone to recognize her signs so that she can communicate with her speaking friends.
Clearly, Mary is unable to create and annotate a dataset and then train a machine learning model for sign recognition.
Instead, Mary (possibly with help from her parents) could describe how each of her signs looks like in natural language.\footnote{We are here considering the problem of isolated sign language recognition for recognizing finger-spelt alphabets given input image, though the more general problem of recognizing signs from a continuous video is harder.}
Given such descriptions from Mary, an expert programmer could setup a pipeline of two foundation models which process an image of Mary signing: first a pose detection model such as MediaPipe for hand shape recognition\footnote{\url{https://developers.google.com/mediapipe/solutions/vision/hand_landmarker}}, whose outputs of coordinates are read and used by an LLM to identify this sign.\footnote{This can be setup as a plugin to ChatGPT or as a pipeline on the cloud by an expert programmer and offered to Mary through her chat app.}
This works well with an example sign with GPT-4.\footnote{See Exhibit in the next page}
If this can be robustly scaled to support various signs, it would be a remarkable change to the current design of machine learning: Mary would create a text-file in her smartphone with her personalized descriptions of signs to build her sign language recognizer!

\newthought{In both the above examples}, we demonstrate how lay users can be enabled to configure and personalize intelligent systems.
The critical difference between this and current machine learning systems is the nature of human data used.
For instance, recommendation algorithms use \textit{low-dimensional data in gigantic volumes}, such as whether a particular user read a particular post amongst the millions of posts and users.
In contrast, our examples suggest usage of \textit{high-dimensional data in very small volumes}, such as the user conveying why he liked a particular post. 
Put simply, LLMs may enable user data in units that correspond to user's thoughts (such as why I liked a video or how I sign a sign) rather than an voluminous bit logs (such as browser history or annotation of many images).
This shift empowers more to heed to Jaron Lanier who challenges ``digital Maoism'' or the idea that collective crowd-sourced intelligence is superior to individual expertise and calls to individuals to reclaim their individuality and critically engage with technology that surrounds them.\cite{lanier2011you}

\begin{figure*}
\includegraphics{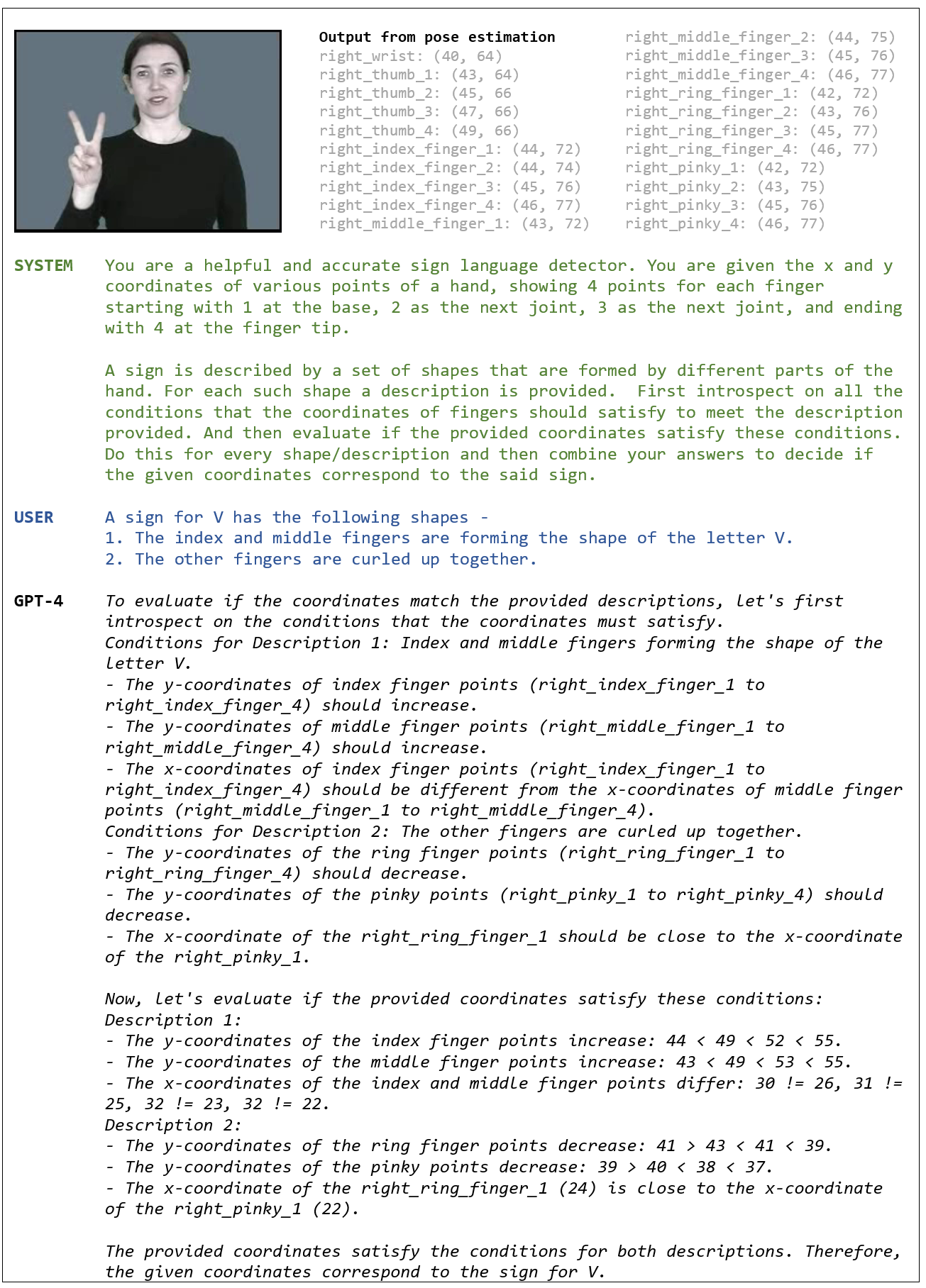}
\caption{Exhibit for Example 6: A Deaf girl provides natural language description for an LLM to recognize signs}
\end{figure*}

\section{How to effectively and rapidly diffuse technology for equity}

There is genuine concern that Large Language Models (LLMs) can lead to higher inequity in society, with few technology companies cornering most of the economic value.
While this may be a distinct possibility, we believe that there is an alternative possibility of using LLMs to drive greater equity with wider access to technologies.
This optimism stems from the above and similar experiments wherein LLMs demonstrate abilities to work across languages, empower people to use a variety of tools as appropriate, and to personalize and create intelligent agents that work for them, all in the well-understood medium of natural language conversation.
Common across all these is greater transfer of agency to people and communities.
To actualize this potential, we suggest a three-point agenda as discussed below.

\subsection{1. Wider Understanding of LLMs}
There is much written about potential risks (including existential) on the one hand, and the huge financial bets and windfalls made by companies on the other hand, due to LLMs.
There is relatively much lesser attention given to how LLM technologies can humanize our experiences with impact across societal strata, especially as a counter to the digital divides and bottlenecks imposed by today's technologies.
This essay, though premature as a scientific contribution, was written to address this lack of public imagination.
Efforts must be made to inform citizens, policy makers, business owners, and technologists of the potential positive use-cases.
This is particularly important because use-cases have to be \textit{discovered} drawing upon the creativity and needs of users across the world.
A case in point is any surprise readers may have experienced in the abilities of LLMs demonstrated in the exhibits in this essay.
Also importantly, these use-cases must inform design of benchmarks that are used by engineers to evolve these LLMs.\footnote{\url{https://github.com/openai/evals}}
For instance, if Jaswinder calling to get his queries answered is an important use-case, then designers of next generation of LLMs must evaluate these models for performance on cross-lingual reasoning.

\subsection{2. Tools with Human Agency}

% - genuine opportunity for communities to build tools for themselves (cite Oslon) - need PbyC
% - need a whole new library of low-code tools that work on top of LLMs
% - need new architectures that combine human infrastructure with software infrastructure (co-pilots for entire flows)

Each of the examples we discussed requires production-grade applications wrapping around LLMs to enable people to have robust and effective experiences.
For being successful in creating equity, such applications must have several properties.
First, these applications must be suffused with human agency: for instance, teachers building more personalized content for their students must be able to easily define guardrails and infuse pedagogic design into the LLM tool that the student would use.
These applications also require thoughtful design to be accessible with short learning curves: for instance, a financial advisor application for a domestic help must be cognizant of her limited digital expertise.
Finally, these tools must enable collaboration across users\footnote{The value of community-led innovation is clear from the rapid evolution of image generation AI with user sharing prompts in social networks like Discord}: for instance students in a Deaf school can collaborate on a sign recognition system by sharing their own text files of descriptions.
Indeed, if LLMs solve increasingly complex tasks, then the role of an application developer may shift to have a much larger focus on the user interface design.
This co-invention\cite{bresnahan1996technical} around the core technology of LLMs is critical for large-scale societal impact.
% Researchers in the areas of ICT4D, startups, and product teams within large companies should be incentivized to undertake this co-invention\cite{bresnahan1996technical} around the core technology of LLMs.

\subsection{3. Cross-institutional Capacity}

LLMs are fairly new in public imagination and are evolving exponentially, with large potential effects on society and economy.
And unlike other technologies, they are not well understood scientifically.
Thus, we need cross-institutional capacity cutting across science, government, and technology companies to articulate considered points-of-view, chart directions of improvement, and regulate bad actors.
The success of developing and scaling Digital Public Goods (DPGs), such as the public identity system in India called Aadhaar\cite{raghavan2019india}, is a good exemplar of such cross-institutional effort.
The DPG playbook encourages private innovation on top of open inter-operable protocols which are used in population-scale systems that are backed by governments.
Applying this lens to LLMs, it is important to build open-source LLMs and technologies around them.
The rapid progress by the open-source community by releasing models such as LLaMA, Alpaca, Dolly, Vicuna, MPT, and others is a remarkable force of equity. 
Though these models trail models such as GPT-4 on general purpose tasks, they will likely enable AI innovators to build models of comparable quality for specific domains.
The second key consideration is ensuring that LLMs are widely accessible to all through the availability of sufficient compute. 
Indeed, companies, foundations, and governments need to rethink their technology investment to have a higher concentration of compute power such as GPUs and accelerators that enable training, fine-tuning, and usage of LLMs. 
This tight coupling of LLM training and usage on hardware will put greater emphasis on sovereignty around semiconductor chips, such as EU's proposed Chips Act.\footnote{\url{https://commission.europa.eu/strategy-and-policy/priorities-2019-2024/europe-fit-digital-age/european-chips-act_en}}

\newthought{In sum,} we remain optimistic that as LLMs develop they will enable us with the agency to solve ``real problems of man and society.''\footnote{Indian space scientist Vikram Sarabhai said, '\textit{we must be second to none in the application of advanced technologies to the real problems of man and society.}'} 

\nobibliography{main.bib}
\bibliographystyle{plainnat}

\end{document}